# Single Quantum Emitters in Monolayer Semiconductors

Yu-Ming He[1,2], G. Clark[3], J. R. Schaibley[4], Yu He[1,2], M.-C. Chen[1,2], Y.-J. Wei[1,2], X. Ding[1,2]

Qiang Zhang[1,2], Wang Yao[5], Xiaodong Xu[3,4], Chao-Yang Lu[1,2], and Jian-Wei Pan[1,2]

[1] Hefei National Laboratory for Physical Sciences at Microscale and Department of Modern Physics, University of Science and Technology of China,

[2] CAS Centre for Excellence and Synergetic Innovation Centre in Quantum Information and Quantum Physics, Hefei, Anhui 230026, China,

[3] Department of Material Science and Engineering, University of Washington, Seattle, Washington 98195, USA,

[4] Department of Physics, University of Washington, Seattle, Washington 98195, USA

[5] Department of Physics and Centre of Theoretical and Computational Physics, The University of Hong Kong, Hong Kong, China,

**Abstract:**

**Single quantum emitters (SQEs) are at the heart of quantum optics[1] and photonic quantum information technologies[2]. To date, all demonstrated solid-state single-photon sources[3-5] are confined in three-dimensional materials. Here, we report a new class of SQEs based on excitons that are spatially localized by defects in two-dimensional tungsten-diselenide monolayers. The optical emission from these SQEs shows narrow linewidths of ~130 $\mu eV$, about two orders of magnitude smaller than that of delocalized valley excitons[6]. Second-order correlation measurements reveal strong photon anti-bunching, unambiguously establishing the single photon nature of the emission[7]. The SQE emission shows two non-degenerate transitions, which are cross-linearly polarized. We assign this fine structure to two excitonic eigen-modes whose degeneracy is lifted by a large ~0.71 meV coupling, likely due to the electron-hole exchange interaction in presence of anisotropy[8]. Magneto-optical measurements also reveal an exciton g-factor of ~8.7, several times larger than that of delocalized valley excitons[9-12]. In addition to their fundamental importance, establishing new SQEs in 2D quantum materials could give rise to practical advantages in quantum information processing, such as efficient photon extraction and high integratability and scalability.**

**Main Text:**

Monolayer semiconductors, such as $MoS_2$ and $WSe_2$, have emerged as a promising platform for both scientific study and technological application due to their exceptional optical and electronic properties. In particular, they provide a novel system to investigate internal degrees of freedom of electrons, i.e. spin and valley pseudospin[13], for potential quantum device applications. A variety of achievements showing control of valley pseudospin have already been demonstrated, including the optical generation of valley polarization[14-17] and coherence[6], the valley Hall effect of valley polarized photo-carriers[18], valley polarized light emitting diodes[19], and the valley selective optical Stark effect[20-21]. All of this progress relies on the optical selection rule associated with the valley pseudospin of delocalized excitons in high quality samples.

Photoluminescence (PL) experiments also reveal excitonic states related to defects, which emit below the energy of the delocalized valley exciton[6,15-17,22]. An interesting question is whether we can utilize these defect-bound excitons for potential quantum technology applications, where the 2D nature of the host lattice could offer potential advantages over traditional solid-state quantum emitters. A first step is to investigate if these defect-bound excitons can serve as single quantum emitters (SQEs), emitting one and only one photon at a time. SQEs are essential for many quantum information processing applications, including quantum key distribution[23], quantum networks[24], and photonic quantum computing[25]. SQEs have been previously realized in a number of solid-state systems, such as single molecules[3], quantum dots[4], and color centers in diamond[5]. Traditional solid-state emitters are typically embedded in three-dimensional materials with high-refractive index, which usually limits the integratability and the photon extraction efficiency. Whereas, the 2D geometry of a SQE confined to an atomically thin material can in principle greatly enhance the photon extraction efficiency, potentially allowing for simplified integration with photonic circuits, and could facilitate strong and controllable external perturbations due to the close proximity of the embedded SQEs.

Here, we report the first observation of photon antibunching from localized SQEs in tungsten-diselenide ($WSe_2$) monolayers. $WSe_2$ monolayers are grown on top of a 300 nm $SiO_2$ on silicon substrate by physical vapor transport[26], a scalable synthesis approach (see Methods). For the optical experiments, the monolayers are held in vacuum inside a cryostat at 4.2 K, where a magnetic field is applied perpendicular to the sample plane (Faraday geometry). Experiments are performed in the reflection geometry where a confocal microscope allows for both laser excitation with a beam focal spot of ~1 μm and collection of the emission (see Methods and Supplementary Fig. S1).

The WSe₂ monolayer is excited using a continuous-wave (cw) laser at a wavelength of 637 nm. Figure 1a shows the emergence of sharp spectral lines, which are red shifted by ~40-100 meV from the PL of the delocalized valley excitons (see right inset of Fig. 1a). With an excitation power of 6 μW, the peak intensity of the sharp lines are ~500 times stronger than the delocalized valley excitons. The left inset of Fig. 1a shows the fine structure of the highest-intensity line (we call SQE1), which is composed of a doublet. The red lines are Lorentzian fits which yield linewidths of ~112 μeV and ~122 μeV (FWHM) and a splitting of 0.68 meV. A statistical histogram on 92 randomly localized emitters from 15 different monolayers is presented in Fig. 1b, yielding linewidths ranging from 58 μeV to 500 μeV, with an average spectral linewidth of 130 μeV, roughly two orders of magnitude smaller than the linewidth of the delocalized exciton PL. The linewidth of these emitters increases dramatically when the temperature is increased (see Supplementary Fig. S2).

The sharp lines are highly spatially localized. Figure 1c illustrates a scanning confocal microscope image of the PL from the emission line centered at 1.7186 eV. The isolated bright spots show that the emission is from localized sites, which are likely excitons bound to atomic defects. These sharp lines show strong saturation behavior as a function of laser power. We investigate the power dependence of the SQE1 peak at 1.7156 eV (left inset of Fig. 1a) as an example. Figure 1d shows the integrated counts as a function of excitation power, demonstrating a pronounced saturation behavior similar to an atom-like two-level system. Under excitation with a 3-ps pulsed laser at 1.78 eV, time-resolved PL of the same emitter is shown in Fig. 1e. It can be well fit to a single exponential decay function with a decay time of $\tau$=1.79±0.02 ns.

Next, we use a Hanbury Brown and Twiss-type interferometer to measure the temporal intensity correlation of the emitted photons from SQE1. Figures 2a and 2b show the photon correlation measurements under cw and pulsed laser excitation, respectively, from which we obtain the second-order correlation at zero delay of $g^2(0)=0.14±0.04$ for cw excitation, and $g^2(0)=0.21±0.06$ for pulsed excitation (see Methods). Both $g^2(0)$ values are well below the threshold of 0.5, which unambiguously proves that it is a SQE.

Many SQEs exhibit several physical properties which imply that they are from the same type of atomic defects. Figure 3a shows a PL intensity plot of five SQEs as a function of polarization detection angle and photon energy. Within the laser spot, five pairs of cross-linearly polarized spectral doublets are visible. Three line cuts extracted from the dashed box in Fig. 3a are plotted in Fig. 3c, where blue, red, and black represent horizontally (H), 45 degree (+), and vertically (V) polarized detection, respectively. We label this emitter SQE2. A near-unity degree of linear polarization is observed, limited predominantly by the detection noise. Interestingly, by applying a magnetic field perpendicular to the monolayer surface, we observe a gradual change from linear polarization to circular polarization. At 5.5 T, the doublet becomes completely cross-circularly polarized, illustrated in Fig. 3e. We attribute larger PL intensity of the low energy state to thermalization of higher energy state population.

The doublet's emission polarization remarkably resembles that of single neutral excitons localized in self-assembled III-V quantum dots[8,27]. There, the electron-hole exchange interaction in the presence of in-plane anisotropy hybridizes the left-circularly ($\sigma^+$) and right-circularly ($\sigma^-$) polarized exciton states, leading to linearly polarized fine structure doublet. By applying a magnetic field in the Faraday geometry, the Zeeman interaction could overcome the exchange energy, resulting in the recovery of circularly polarized transitions[27]. This explanation from III-V quantum dots is highly consistent with our observations on monolayer SQEs. Figure 3d depicts a zero-magnetic field energy level diagram based on this model. The statistics of the exchange splitting from 16 different SQEs are shown in Fig. 3b. We observe an average exchange splitting of $0.71 \pm 0.04$ meV, roughly an order of magnitude larger than that of typical III-V quantum dots[8,27]. Moreover, these splittings show a significantly higher level of uniformity than III-V quantum dots[27].

This picture of the exchange splitting is further supported by the magnetic field dependence of the emission polarization in Faraday geometry. Under a magnetic field of 5.5 T, the Zeeman splitting of SQE2 exciton is ~2.85 meV (see Fig. 4), which overwhelms the measured exchange splitting. Therefore, the hybridization between left- and right- polarized exciton states is suppressed and circularly polarized emission is restored (Fig. 3f). This magnetic field dependence of emission polarization further suggests that a possible origin of the defect emission is due to the confinement of valley excitons at defects with broken rotational symmetry. Given the $C_3$ rotational symmetry of the 2D $WSe_2$ crystal, delocalized excitons in the two opposite valleys are coupled to left- and right- circularly polarized fields. When the exciton is subject to confinement that lacks the $C_3$ rotational symmetry, the intervalley electron-hole exchange interaction can couple the left- and right-circularly polarized valley excitons, resulting in new eigenstates which couple to linearly polarized fields along two orthogonal axes[28].

Figure 4a shows a PL intensity plot as a function of applied magnetic field (0 T to 5.5 T), which shows the evolution of the five doublets in Fig. 3a. The extracted energy positions of the SQE2 doublet, and its Zeeman splitting are plotted as a function of magnetic field in Fig. 4b and 4c, respectively. We note that a random yet synchronized spectral wandering of the doublet occurred on the measurement time scale and can be observed in Fig. 4b. The spectral wandering is absent in the extracted Zeeman splitting, i.e. the difference between the two emission lines (Fig. 4c), and provides further evidence that the doublet emission is from the same SQE. The magnetic field dependent energy splitting can be well fit to $E = \sqrt{(g\mu_B B)^2 + \delta_1^2}$, where $\delta_1$ is the exchange splitting as shown in Fig. 3b, $B$ is the magnetic field, and $\mu_B$ is the Bohr magneton. We extract a $g$-factor of $8.897 \pm 0.003$ for SQE2. For 16 random samples, 15 of them show a relatively uniform distribution ranging from $8.16 \pm 0.05$ to $9.45 \pm 0.10$ (see Supplementary Figs. S3-S4), whereas one shows a relatively small $g$ factor of $5.97 \pm 0.04$. These $g$-factors are larger than excitons in self-assembled InGaAs quantum dots[27] and delocalized valley excitons[9-12] in monolayer WSe$_2$. We note that we also measured another type of SQE which does not show the fine structure or Zeeman splitting (see Supplementary Figs. S5-S6). The origin of these emitters are not known and will be interesting for future study.

In summary, we have reported a new type of SQE native to a 2D quantum material. These localized SQEs emit single photons with a narrow-linewidth and are composed of a linearly polarized doublet. Polarization-resolved and magneto-optical studies reveal a large zero-field splitting of ~0.71 meV and an exciton $g$-factor of ~8.7. These results suggest that the SQEs are composed of neutral excitons trapped at anisotropic confining potentials from defects in the monolayer. An open challenge is to identify the physical structure of these SQEs, and the origin of the exceptionally large $g$-factor. Our results may open up a new avenue towards photonic quantum devices based on 2D semiconductors. We expect interesting applications including the integration of monolayer SQEs with nanocavities, such as photonic crystals or plasmonics, to control the single photon emission rate. We are also encouraged by the possibility of using scanning tunneling microscopy[29] techniques which may facilitate deterministic engineering of SQE arrays for scalable quantum technology applications.

During preparation of the manuscript, we became aware of similar work posted on Arxiv[30].

**Methods:**

WSe$_2$ Growth: WSe$_2$ monolayers are grown by physical vapor transport using powdered WSe$_2$ as precursor material. Source material (30 mg) in an alumina crucible is placed in the hot zone of a 1 in. horizontal tube furnace, and the SiO$_2$ substrate is placed downstream in a cooler zone (750-850 °C) at the edge of the furnace. The furnace is heated to 950 °C at a rate of 35 °C/min and remains there for a duration of 5 min before cooling to room temperature naturally. The system is held under vacuum. A flow of 80 sccm argon and 25 sccm hydrogen is introduced as carrier gas during the 5 min growth period. Details can be found in Ref. 26.

Optical Measurements: A diagram of our experimental setup is shown in Fig. S1. The sample were optically excited nonresonantly using a continuous-wave tunable laser tuned to 637 nm with a bandwidth of ~0.05 pm. In the excitation arm, a 635-645nm band-pass filter is used to filter out the unwanted Raman scattering from the optical fibre. In the collection arm, residual pump laser light was filtered out using a 660 nm long-pass filter. A spectrometer with a liquid-nitrogen-cooled CCD was used to analyse the PL. To determine the emission spectral linewidth distribution in Fig. 1b, we use a 1800-grating monochromator with a spectral resolution of ~20 μeV. Hanbury Brown and Twiss photon-correlation measurements are performed using two single photon counting avalanche photodiodes. A record of coincidence events is kept to build up a time-delay histogram, as shown in Fig. 2. The cw data in Fig. 2a is fit with a double exponential decay function. The pulsed $g^2(0)$ value is calculated from the integrated photon counts (12.5 ns time windows) in the time-zero delay peak divided by the average of the adjacent six peaks. The error on $g^2(0)$, which denotes one standard deviation, is deduced from the propagated Poissonian counting statistics of the raw detection events.

**Acknowledgement**: This work is supported by the National Natural Science Foundation of China, the Chinese Academy of Sciences, the National Fundamental Research Program. The work at U. Washington is supported by AFOSR (FA9550-14-1-0277). GC is partially supported by the State of Washington through the University of Washington Clean Energy Institute. XX thanks the support from Cottrell Scholar Award. WY is supported by the Croucher Foundation (Croucher Innovation Award), and the RGC of Hong Kong (HKU705513P, HKU9/CRF/13G).

## Competing Financial Interests

The authors declare no competing financial interests.

**Figures:**

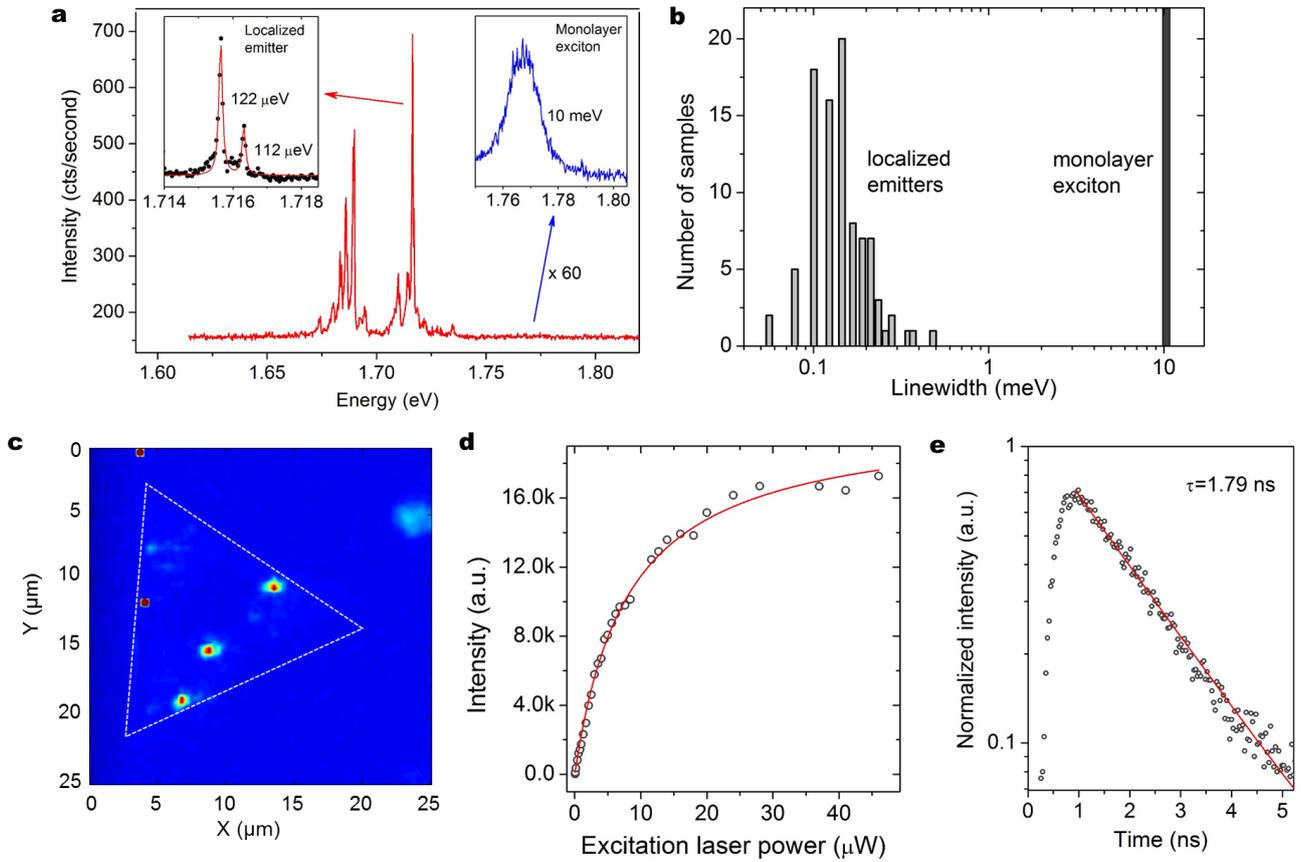

**Figure 1 | Emergence of red-shifted narrow spectral lines in monolayer WSe₂. a,** PL spectrum of localized emitters. The left inset is a high-resolution spectrum of the highest intensity peak (SQE1) with a spectral resolution of ~30 μeV using a 1200-grating monochromator. The right inset is a zoom-in of the monolayer valley exciton emission, integrated for 60 s. The emission of the localized emitters exhibits a red-shift and much sharper spectral lines. **b,** A histogram comparison of the linewidth between the emission from the delocalized valley exciton and 92 localized emitters, grouped in 23.3 μeV bins. **c,** PL intensity map of a narrow emission at 1.7186 eV over a 25 μm × 25 μm area. The dashed triangle indicates the position of the monolayer. **d,** The integrated counts of the photon emission from SQE1 showing saturation behavior with increasing laser power. The red line is a guide to eye. **e,** Time-resolved PL of SQE1, showing a single exponential with a decay time of 1.79±0.02 ns.

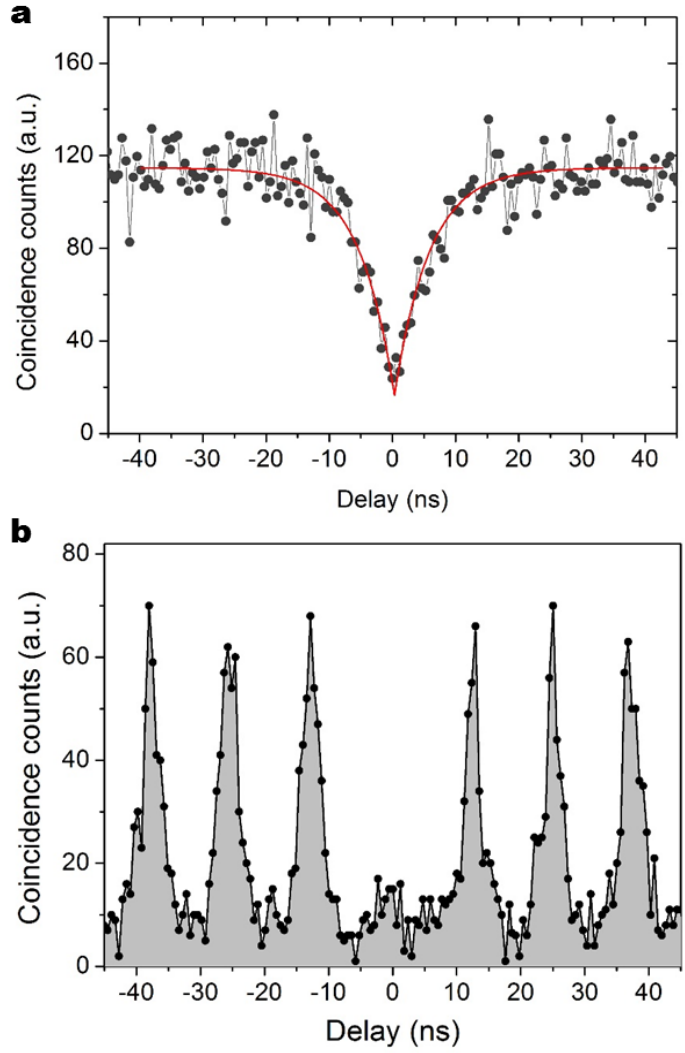

**Figure 2 | Observation of photon antibunching**. **a,** Second-order correlation measurement of the PL from SQE1 under 6.8 μW continuous-wave laser excitation at 637 nm. The red line is a fit to the data, from which we extract $g^2(0) = 0.14 \pm 0.04$. **b,** Intensity correlation measurement under an 9.5 μW (average) pulsed excitation at 696.3 nm with a repetition rate of 82 MHz, and a pulse width of ~3 ps, revealing a $g^2(0) = 0.21 \pm 0.06$.

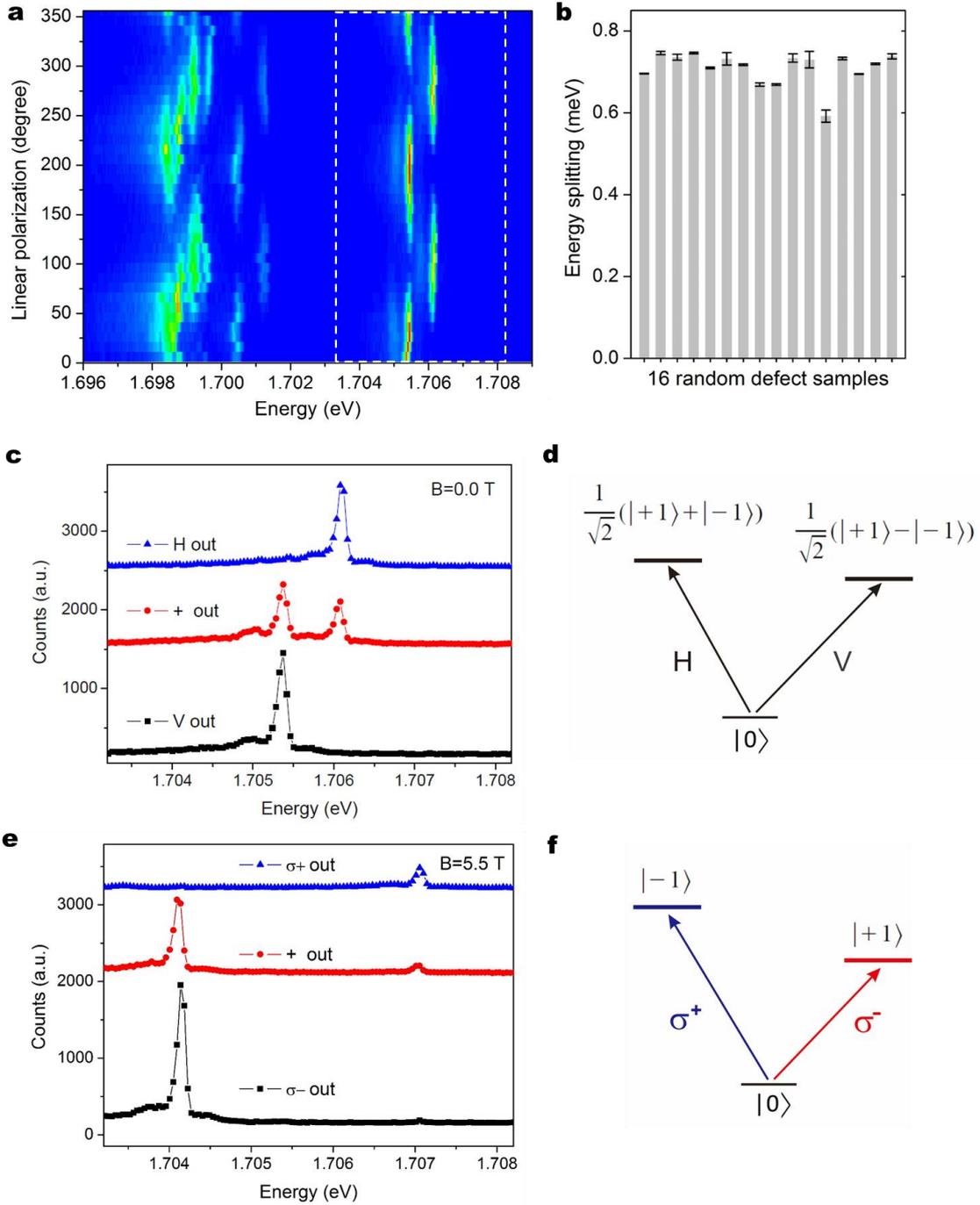

**Figure 3 | Polarization-resolved PL from SQEs**. **a,** PL intensity plot of SQEs as a function of polarization detection angle and photon energy. Five pairs of cross-linearly polarized spectral doublets are observed at: 1.6984-1.6692 eV, 1.6988-1.6996 eV, 1.7005-1.7012 eV, 1.7050-1.7058 eV, and 1.7054-1.7061 eV (SQE2). **b,** The fine structure splitting measured on 16 different SQEs varies from 0.59 meV to 0.75 meV. **c,** Polarization-resolved PL spectra of SQE2 at zero magnetic field showing a cross-linearly polarized doublet. **d,** Zero-field energy level diagram with linearly polarized optical selection rules. **e,** At a magnetic field of 5.5 T, the doublet becomes cross-circularly polarized, and **f,** shows its energy level diagram. H, V, $\sigma^+$, $\sigma^-$, and + denote horizontal, vertical, left-circular, right-circular, and +45 linear polarization, respectively.

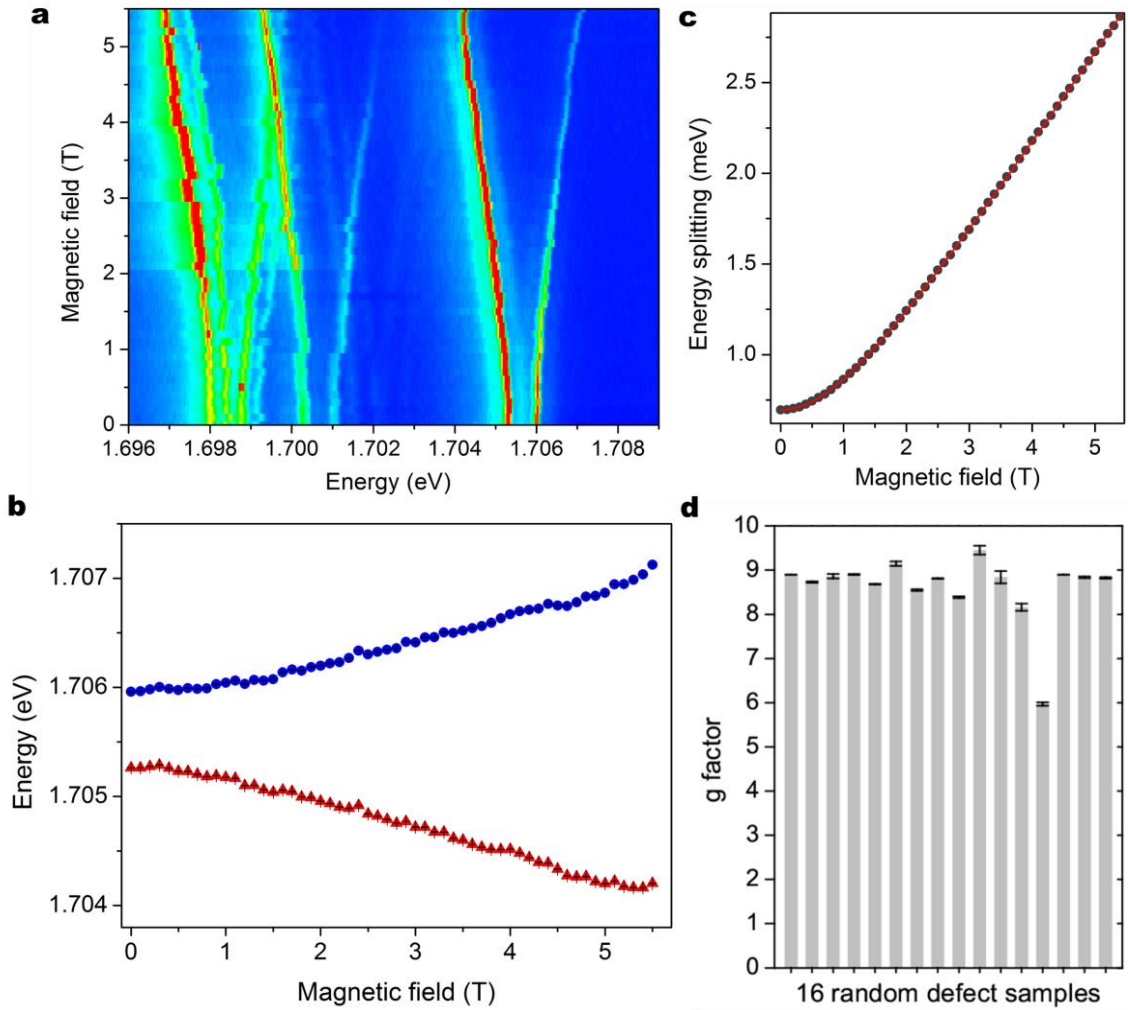

**Figure 4 | Magneto-optical measurement of SQEs**. **a,** PL intensity plot of five doublets, shown in Fig. 3a, as a function of applied magnetic field perpendicular to the sample. Spectra were taken every 0.1 T. **b,** Extracted central energies of the SQE2 doublet. **c,** Extracted energy splitting between the SQE2 doublet in b as a function of magnetic field. **d,** The exciton g-factor of 16 different emitters.

# Single Quantum Emitters in Monolayer Semiconductors

Yu-Ming He, G. Clark, J. R. Schaibley, Yu He, M.-C. Chen, Y.-J. Wei, X. Ding, Qiang Zhang,

Wang Yao, Xiaodong Xu, Chao-Yang Lu, and Jian-Wei Pan

## Supplementary Figures

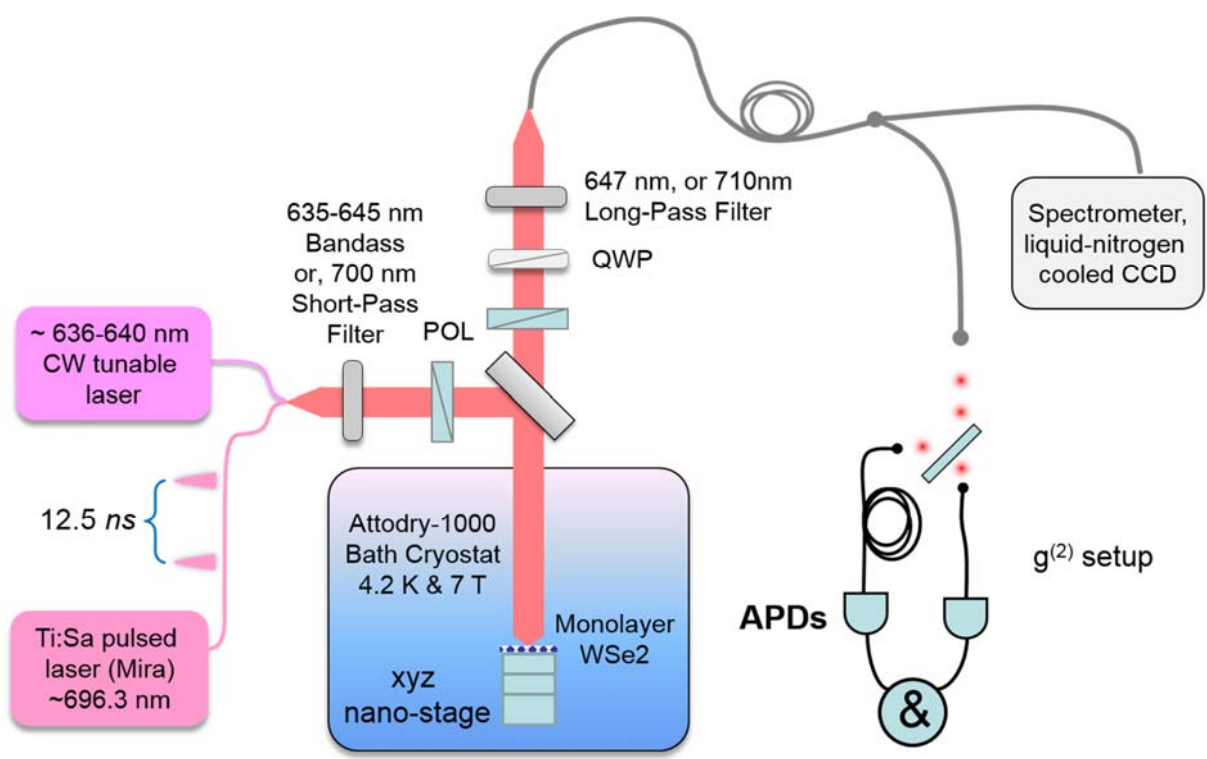

**Figure S1 | An illustration of the optical spectroscopy setup.** The monolayer WSe2 sample is housed in an Attodry 1000 bath cryostat at 4.2 K with ultralow vibration (<1 nm) which allows us to focus on the same single emitters for long-term (~months) study. A superconducting magnet can generate variable magnetic field up to 7 T. The sample was positioned relative to the objective using piezoelectric 'slip-stick' positioners. An aspheric objective lens with NA=0.82 was placed inside the cryostat to focus both excitation laser and collect the emitted single-photon fluorescence. For cw excitation, we use a tunable narrowband laser at around 637nm. A band-pass filter (635-

645 nm) is used in the excitation arm to filter out the unwanted Raman scattering in the optical fibre. In the output arm, we use a 647 nm long pass filter to suppress the excitation laser background. For pulsed excitation, we use a mode-locked Ti:Sapphire laser with a pulse duration of ~3ps, repetition rate of 82 MHz, and wavelength of ~696.3 nm (with a higher energy than the both the delocalized valley exciton and the SQEs). A 710 nm long pass filter is used to filter out the excitation laser scattering. For spectral analysis, the emitted photons are sent to a spectrometer with three different choices of gratings (300, 1200 and 1800 gratings) and detected by a liquid-nitrogen-cooled CCD. For time correlation measurement, we sent the light stream through a 50:50 beam splitter which are then detected by two single photon counting avalanche photodiodes (APDs). The coincidence events is recorded to build up the time-delay histogram ($g^2(\tau)$). POL: polarizers. QWP: quarter wave plate.

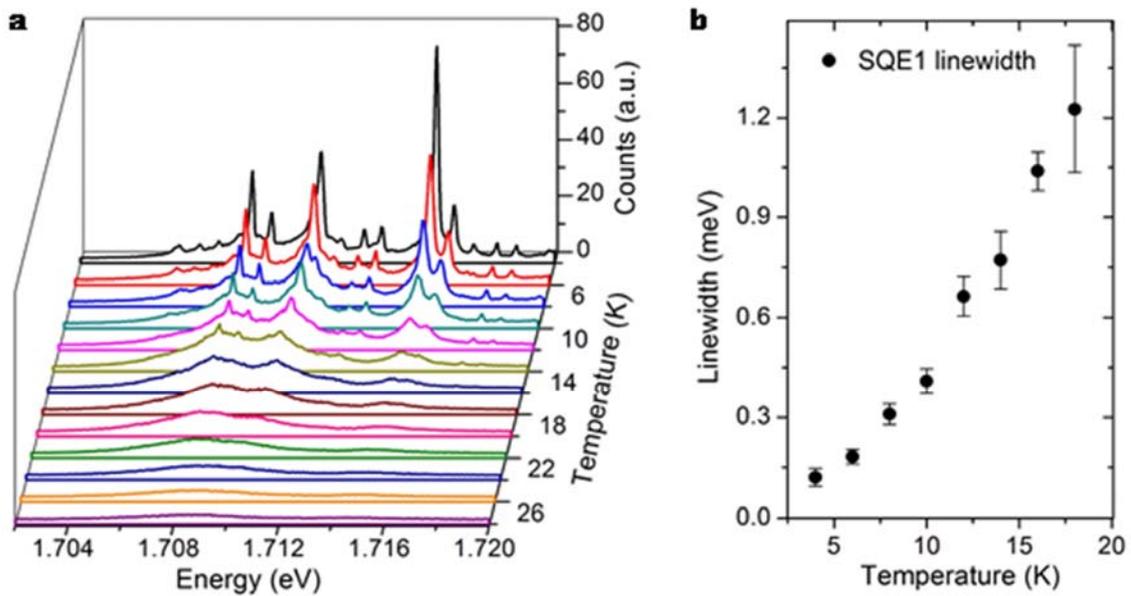

**Figure S2 | Temperature dependence of single quantum emitters (SQE2). a**, Temperature-dependent PL spectra with a 1200-grating spectrometer corresponding to the data in Fig. 1a in the main text. **b**, Extracted average linewidth for the SQE1 doublet.

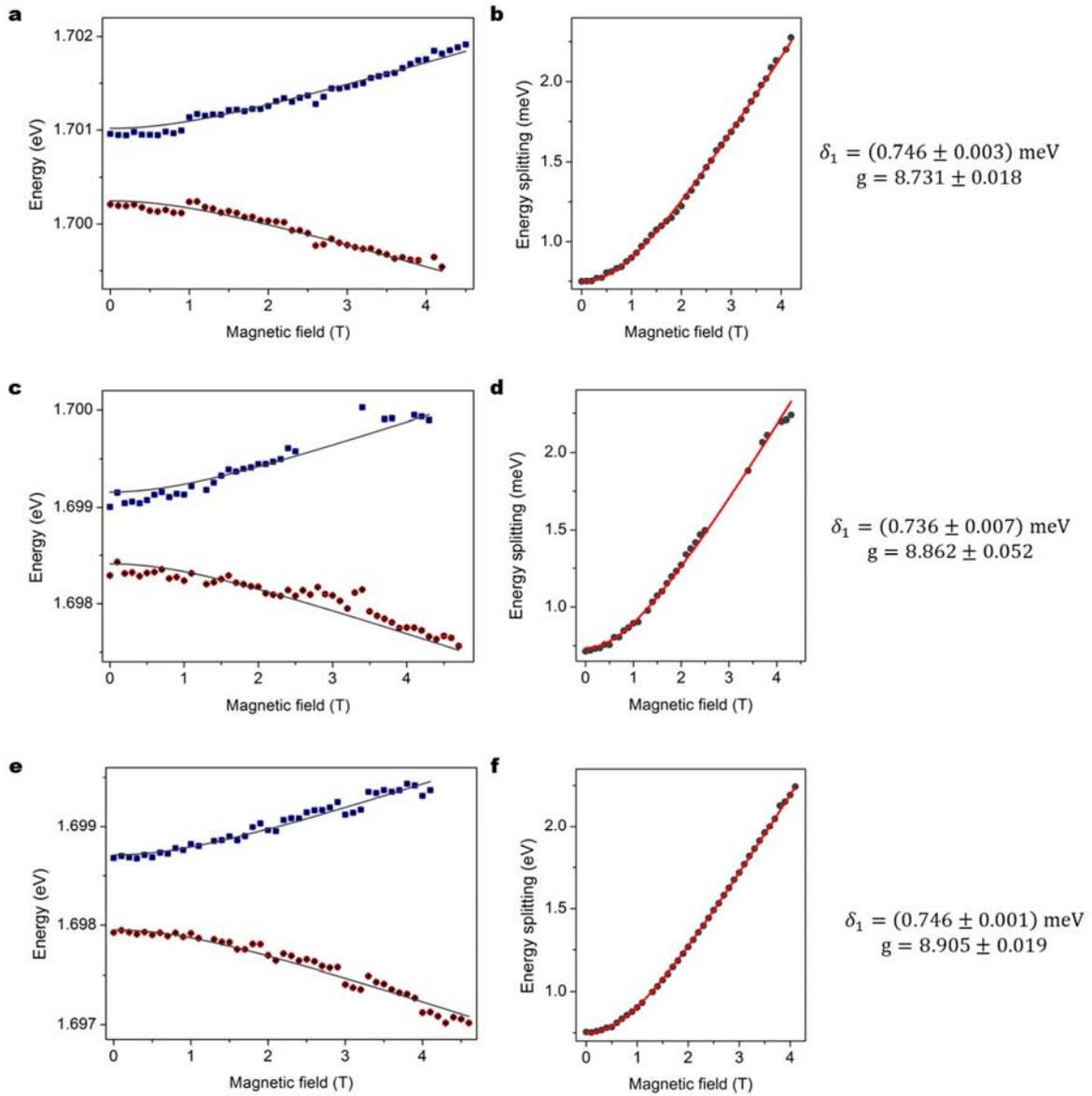

**Figure S3 | Zeeman splitting of the other three doublets in Fig. 4a (in the same monolayer as SQE2). a, c, e,** The extracted energies of the doublet as a function of magnetic field. **b, d, f,** The energy splitting as a function of magnetic field. The fits in **b, d, f,** extract the exchange splitting energy and the g-factor for each emitter, and is listed in the right. Again, we note that a random yet synchronized spectral wandering of each doublet occurred on the measurement time. The spectral wandering is absent in the extracted Zeeman splitting, supporting that each doublet is from the same SQE.

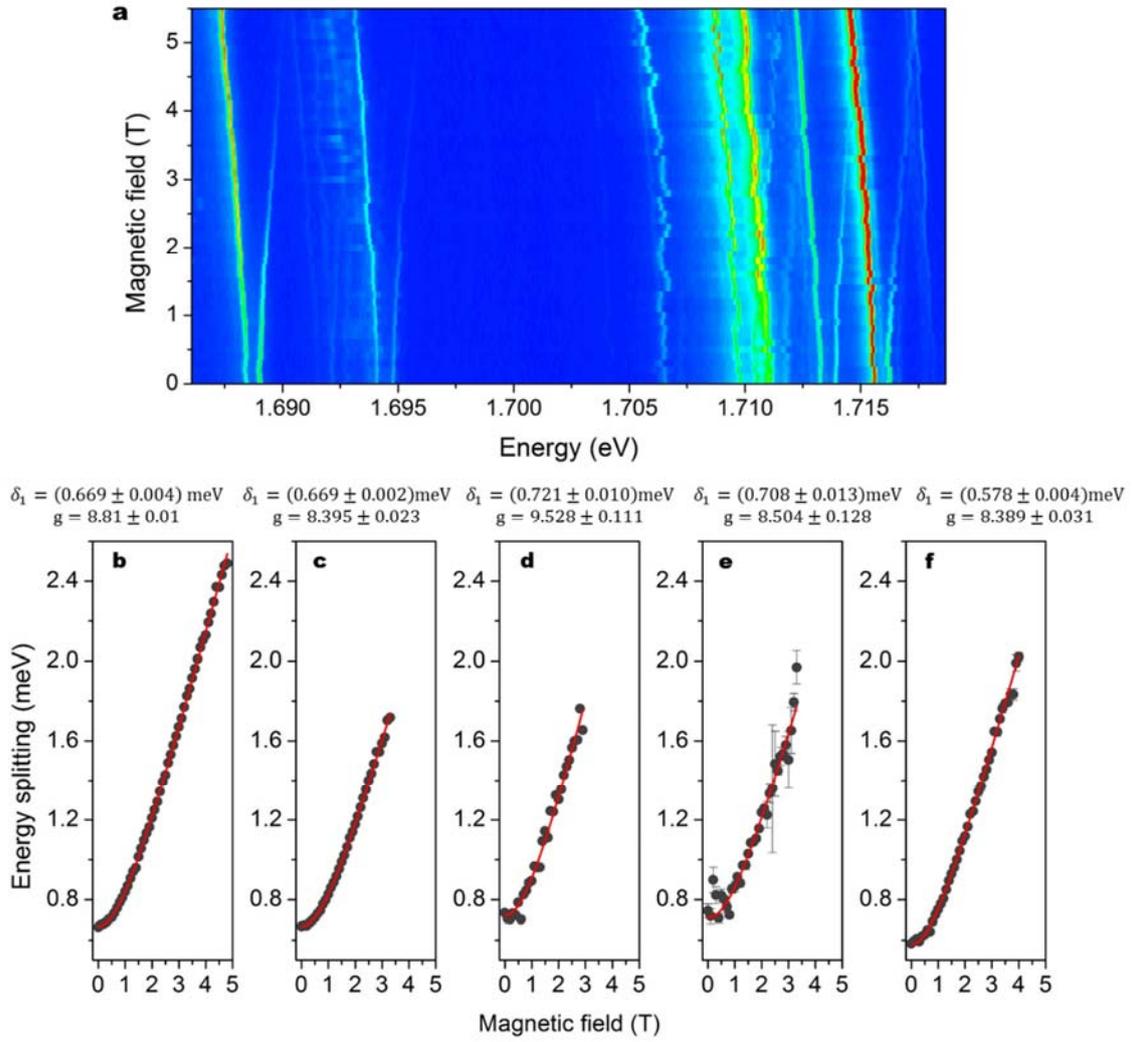

**Figure S4 | The magnetic field dependent photoluminescence of SQE1 and four nearby emitters in the same monolayer. a,** Photoluminescence intensity plot as a function of magnetic field and photon energy. **b – f** show the extracted Zeeman splittings of five doublets, with exchange splitting and g-factor indicated on top of each panel. Red lines are fit according to the equation given in the main text.

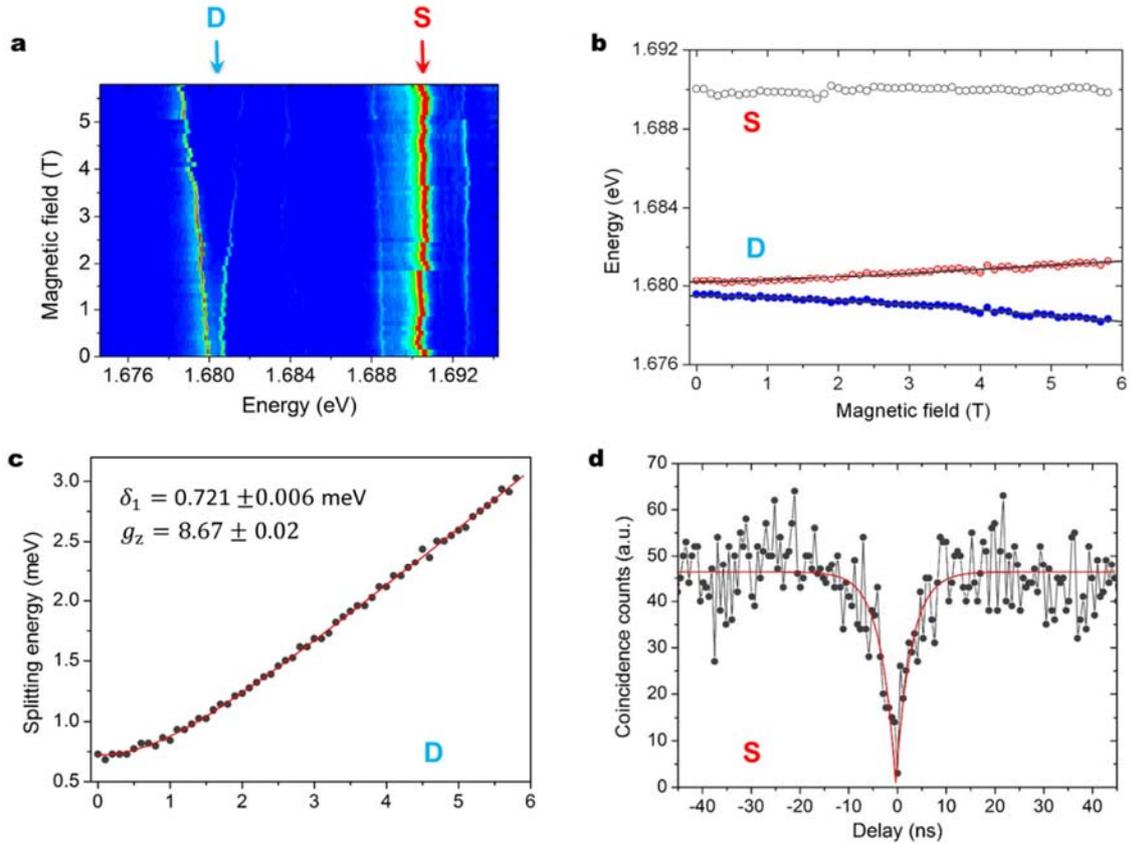

**Figure S5 | Comparison of two types of single quantum emitters**. In our manuscript, we have mainly focused on single quantum emitters (SQEs) which show exchange interaction splitting. We also find with a ~40% probability (37 out of 92 investigated emitters) another type of SQEs that have only a single line and do not split in the presence of a magnetic field. **a**, Photoluminescence intensity plot as a function of magnetic field and photon energy, displaying SQEs with a doublet (labelled as D) and with a single peak (labelled as S). **b**, Extracted peak energies of both S (black hollow circles) and D (red and blue circles). **c**, Zeeman splitting of the D emitter, similar as presented in the main text. The S emitter shows a spectral wandering but no pronounced Zeeman splitting is observed. **d**, Second-order correlation measurement of the S emitter by cw excitation. Strong photon antibunching (0.05) is observed. The S line does not show any splitting or shift, beyond the resolution of our experiments (~30 µeV) and the spectral diffusion for this emitter (FWHM~68 µeV, see Fig. S6b). This might indicate a transition between ground and excited states with the same *g* factors, similar to the chromium-based colour centres in diamond (see Müller, T., *et al.* New. J. Phys 13, 075001 (2011)).

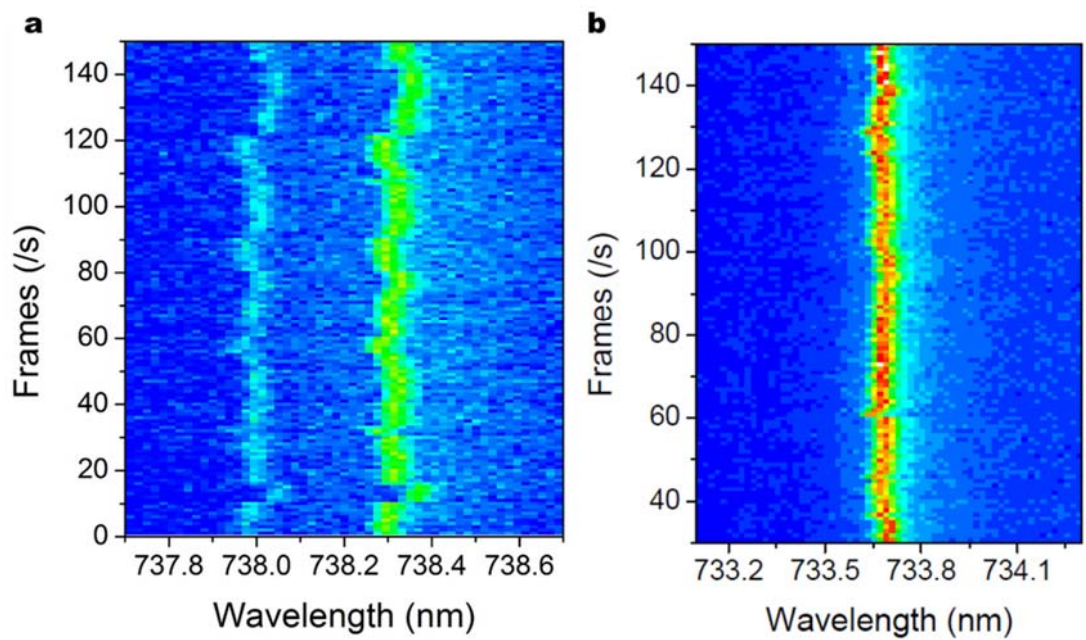

**Figure S6 | Spectral wandering of the SQEs**. Photoluminescence intensity plot of **a**, a cross-linearly polarized doublet, and **b**, a single peak as a function of photon energy and frame number (time). Each frame is integrated for 1s.